\documentclass{aa}
\usepackage{graphics, epsfig}
\begin{document}

%\thesaurus{10.08.1; 12.07.1; 12.04.1;10.19.3}

\title{Gravitational lensing potential reconstruction in quadruply imaged systems}

\author{V. F. Cardone\inst{1}
        \and S. Capozziello\inst{1}
        \and V. Re\inst{1}
    \and E. Piedipalumbo\inst{2}}

\offprints{V.F. Cardone, \email{winny@na.infn.it}}

\institute{Dipartimento di Fisica ``E.R. Caianiello'', Universit{\`a} di Salerno and INFN, Sezione di Napoli, Gruppo Collegato di Salerno, Via S. Allende, 84081 - Baronissi
(Salerno), Italy \and Dipartimento di Scienze Fisiche, Universit{\`a} di Napoli and INFN, Sezione di Napoli, Complesso Universitario di Monte S. Angelo, Via Cinthia, Edificio N - 80126 Napoli, Italy}

\date{Receveid / Accepted }

\abstract{We develop a semi-analytical method to reconstruct the lensing potential in quadruply imaged gravitational lens systems. Assuming that the potential belongs to a broad class of boxy non\,-\,elliptical models, we show how it is possible to write down a system of equations which can be numerically solved to recover the potential parameters directly from image positions and using physical constraints. We also describe a code developed to search for solutions of the system previously found and test it on simulated cases. Finally, we apply the method to the quadruple lens PG1115+080 which allows us to get also an estimate of the Hubble constant $H_0$ from the measured time delay as $H_0 = 56_{-11}^{+12} \ {\rm km \ s^{-1} \ Mpc^{-1}}$.
}

\titlerunning{Lensing potential reconstruction}

\maketitle

\section{Introduction}

Since the very beginning with the discovery of the first double lens QSO0957 +561 (\cite{WCW79}), gravitational lensing has found many cosmological applications (see, e.g, \cite{NB98} and references therein). Measurements of time delays in gravitational lenses can be used to give constraints to the value of the Hubble constant $H_0$ (\cite{R64}), avoiding the usual calibration problems that plague local distance estimators (see, e.g., \cite{F00}). In a gravitational lensing system, the ray trajectories of the multiple images have different paths and pass through different parts of the gravitational potential, so that the light travel time is different for each image. In cosmology, it is easy to show that the light travel time is inversely proportional to $H_0$, so that by the observed time delays between images it is possible to deduce the value of the Hubble constant. However, being the distance $D = cz/H_0$ (for $z < 1$), the gravitational lensing potential has a key role in the determination of $D$. Unfortunately, application of this technique has been slowed by the difficulties of determining time delays (see, e.g., the remarks in \cite{S00}) and of finding a unique model of the gravitational lensing potential since there  are often sequences of lens models which fit the same image positions.

Up to now, about fifty multiply imaged quasars have been discovered (see \cite{CASTLE} for a detailed list) with fifteen quadruple systems. In all cases, the lensing system consists of a quasar acting as source and a galaxy (or a group of galaxies) acting as lens. These systems furnish a good tool to probe the potential of the lens galaxies. The advantage with respect to the traditional stellar dynamics method is that, by the image positions, we can measure the shape and the mass of the dark halo well beyond the visible edge of a faraway lens galaxy. In this sense, quadruply imaged sytems are better constrained than double quasars since the lens models have to fit more image positions and the time delays between any pair of images. On the other hand, the possible constraints from flux ratios have to be used with more caution due to the possible systematic influence of microlensing effects (\cite{CR79}). In this paper we concentrate only on quadruply imaged systems.

Starting from the image positions, a significant amount of numerical computation is usually required to derive the parameters of  lensing potential (see, e.g., \cite{SEF}). The degeneracy of the results is often not fully explored since a large parameter space should be analyzed using a huge amount of computer time. Previously, several authors have often restricted their studies to isothermal or power\,-\,law spherical models (\cite{EW98}), elliptical models (\cite{WM97}) and other simple models (Kassiola \& Kovner, 1993, 1995) with or without external shear. The fully general non-parametric method, e.g. the {\it pixellated lens method} (\cite{WS00}), is very powerful to derive the range of degeneracy in the lens models, but it involves significant amount of numerical computation and does not provide a clear insight to the relations between the characteristic parameters of the lens. For these reasons, it is still desiderable to find analytical, yet general potentials, which allow a quick exploration of the parameter space. First steps in this direction are the analytical studies of elliptical potentials, with and without external shears due to Witt (1996) and Witt \& Mao (1998, 2000). Witt, Mao \& Keeton (2000) have shown that the estimate of the Hubble constant from the time delay does not depend on the angular shape of the potential provided that the model is isothermal and with no external shear. Beside, they have also shown how it is possible to analitically estimate the error in the estimate of $H_0$ due to assuming incorrectly an isothermal model with no external shear. A further step has been done by Zhao \& Pronk (2001) who studied a broad class of analytical models with non\,-\,elliptical and semi\,-\,power\,-\,law radial profiles. These authors have shown how to reconstruct the lens shape and the radial profile parameters directly from image positions with a semi-analytical method, but they still have to assign {\it a priori} the ``boxiness'' parameter so that their method still needs some improvements. The aim of our work is to develop a semi-analytical method to recover the lensing potential parameters and also the source positions using only image positions and some constraints coming from physical considerations. As a first step, we consider a wide class of lensing potentials similar to the one studied by Zhao \& Pronk (2000) and neglect the external shear, which will be considered elsewhere.

The outline of this paper is as follows. In Sect. 2 we show how it is possible to write down a set of equations to recover the source coordinates and all the lensing potential parameters for a broad class of boxy non elliptical potentials using as input only the image postions. Solving this system requires numerical methods and that is why we have developped a code which we describe in Sect. 3 where we also test the method itself on simulated cases. Sect. 4 is devoted to the application to a real quadruple system, PG1115+080\,: we reconstruct the lensing potential and find the source coordinates and then use these results to estimate the Hubble constant from the measured time delay between images $B$ and $C$. Finally, discussion of the results and future improvements and applications of the method are presented in Sect.5. 

\section{Lensing potential parameters from image positions}

Let us choose a rectangular coordinate system $(x,y)$ with origin on the lens galaxy centre and axes pointing towards West and North respectively. Let $(r, \theta)$ be the polar coordinates being $\theta$ the position angle measured
counterclockwise from North; then we have the following coordinates transformation\,:

\begin{displaymath}
x = r \sin{\theta} \ , \ \ \ \ y = r \cos{\theta} \ .
\end{displaymath}

The time delay of a light ray deflected by the galaxy lensing effect is given by\,:

\begin{displaymath}
\Delta t (r, \theta) = h^{-1} \tau_{100} \ \times 
\end{displaymath}
\begin{equation}
\ \ \ \ \ \ \ \ \ \ \ \ \
\left [ \frac{1}{2} r^2 - r r_s \cos{(\theta - \theta_s)} + \frac{1}{2}
r_s^2 - \psi(r, \theta) \right ] \label{eq: timedelaygen}
\end{equation}

being $(r, \theta)$ the image position, $(r_s, \theta_s)$ the unknown source position and $\psi(r, \theta)$ the
lensing potential. In Eq.(\ref{eq: timedelaygen}), $h$ is the Hubble constant $H_0$ in units of 100 km/s/Mpc
(i.e. $H_0 = 100 h$\,km/s/Mpc), while $\tau_{100}$ is a typical time delay estimated for a given set of
cosmological parameters $(\Omega_m, \Omega_{\Lambda}, \Omega_{k})$ assuming $H_0 = 100$\,km/s/Mpc. We have\,:

\begin{displaymath}
\Omega_m = \frac{8 \pi G \rho_{crit}}{3 H_0} \ , \ \ \ \Omega_{\Lambda} = \frac{\Lambda c^2}{3 H_0^2} \ , 
\ \ \ \Omega_k = -\frac{k c^2}{a^2 H_0^2}
\end{displaymath}

which are the density parameters relative to matter, cosmological constant and spatial curvature respectively. The
typical time delay is defined as\,:

\begin{equation}
\tau_{100} = \left(\frac{D_{OL} D_{OS}}{D_{LS}}\right) \frac{(1+z_L)}{c} \label{eq: taucento}
\end{equation}

with the usual meaning for the angular diameter distances $D_{OL}, D_{OS}, D_{LS}$; $z_L$ is the redshift of the
lens.

According to the Fermat principle, the images lie at the minima of $\Delta t$, so that the lens equations may be
simply obtained minimizing $\Delta t$. We get\,:

\begin{equation}
\frac{\partial}{\partial r}\Delta t = 0 \iff r - r_s \cos{(\theta - \theta_s)} = 
\frac{\partial \psi}{\partial r} \ , 
\label{eq: lenseqa}
\end{equation}

\begin{equation}
\frac{1}{r} \frac{\partial}{\partial \theta}\Delta t = 0 \iff r_s \sin{(\theta - \theta_s)} = \frac{1}{r}\frac{\partial\psi}{\partial \theta} \ . 
\label{eq: lenseqb}
\end{equation}

The lensing potential $\psi(r, \theta)$ is usually made out by the sum of two terms; the first one is connected
to the lens galaxy surface mass density $\Sigma(r, \theta)$ through the relation\,:

\begin{equation}
\nabla^2 \psi = 2 \frac{\Sigma(r, \theta)}{\Sigma_{crit}} = 2 \kappa \label{eq: psikappa}
\end{equation}

being $\Sigma_{crit} = (c^2/4\pi G) (D_{OS}/D_{OL}D_{LS})$ and $\kappa$ the dimensionless surface mass density.

The second contribution comes from the so\,-\,called external shear which we neglect in this analysis\footnote{We
will return on this topic in the conclusions and further again in a following paper.}. In studying quadruple
systems, one may follow two different, but similar, ways. On the one hand, one may give $\Sigma(r, \theta)$ and
then solve Eq.(\ref{eq: psikappa}) for $\psi(r, \theta)$. This ensures that the lens galaxy model has a physical
meaning, but it is not always possible to find analytical solutions of Eq.(\ref{eq: psikappa}). This is a severe
problem since one does not know {\it a priori} the exact values of the model parameters and so cannot integrate
numerically Eq.(\ref{eq: psikappa}). On the other hand, one may directly give $\psi(r, \theta)$, determine the
lensing potential parameters and then solve Eq.(\ref{eq: psikappa}) to get $\Sigma(r, \theta)$ (assuming that one
also knows the redshift of lens and images  to evaluate $\Sigma_{crit}$). At the end, one has to check if the
derived model is a physical one. We follow this second approach since we are mainly interested to reconstruct the
lensing potential directly from observed image positions using a semi-analytical approach. To this end we have
first to choose an expression (as general as possible) for $\psi(r, \theta)$. We limit our attention to the class
of  potentials

\begin{equation}
\psi(r, \theta) = r^{\alpha} \left [ 1 - \delta \cos{(2\theta - 2\theta_p)} \right ]^{\beta} = 
r^{\alpha} \cal{F}(\theta) 
\label{eq: psiour}
\end{equation}

with the constraint $\delta < 1$. It is evident the factorization of the models which are physically well
motivated for the following reasons. In Eq.(\ref{eq: psiour}), $\theta_p$ is the position angle of the main axis
of the lensing potential which may be also misaligned with respect to the main axis of the lens model due to the
possible effect of the external shear or to the influence of other galaxies of the same group. For these reasons,
it is better to not fix $\theta_p$ equal to the value observed for the lens galaxy so that it should be treated
as an unknown parameter of the lensing potential. In Eq.(\ref{eq: psiour}), $\beta$ is a boxiness parameter and it
is defined such that $\cal{F}(\theta)$ reduces to the usual elliptical form when $\beta = 1/2$ (\cite{WM97}),
whilst it gives the simple models of Kassiola \& Kovner (1995) for $\alpha = \beta = 1$. Redefining in the
correct way the parameters, Eq.(\ref{eq: psiour}) may describe also the class of models studied by Zhao \& Pronk
(2001) to model the quadruple system PG1115 + 080. The same authors suggest that $| \beta | < 1/2$ in order to
have physical mass\,-\,radius relation and show that the flattening of the surface mass density of the model
obtained solving Eq.(\ref{eq: psikappa}) for the potential (\ref{eq: psiour}) is given by\,:

\begin{equation}
q_{\kappa} = \left ( \frac{1-\delta}{1+\delta} \right )^{\frac{1-\beta}{2-\alpha}} \left [ \frac{1 - \left (
\frac{4 \beta}{\alpha^2} - 1 \right ) \delta} {1 + \left ( \frac{4 \beta}{\alpha^2} - 1 \right ) \delta} \right
]^{\frac{1}{2 - \alpha}} \ . 
\label{eq: qkappa}
\end{equation}

We have checked that for many values of $\alpha$ and $\beta$, in the physically interesting range, $q_{\kappa} \le
1$ (i.e. the galaxy is oblate or spherical) only if $\delta \le 0$ so that our choice to consider only models
with $\delta < 1$ does not exclude physically interesting cases. It is also easy to see that the surface mass
density $\Sigma(r, \theta)$ scales with the radius as $r^{\alpha -2}$, so that we may impose the constraint
$\alpha < 2$ in order that $\Sigma(r, \theta)$ is a monotonically decreasing function of the radius.

Let us now intsert Eq.(\ref{eq: psiour}) into Eqs.(\ref{eq: lenseqa}) and (\ref{eq: lenseqb}); the lens
equations are thus\,:

\begin{equation}
r - r_s \cos{(\theta - \theta_s)} = \alpha r^{\alpha - 1} \cal{F}(\theta) \ , \label{eq: lenseqoura}
\end{equation}

\begin{equation}
r_s \sin{(\theta - \theta_s)} = r^{\alpha - 1} \frac{d\cal{F}}{d\theta} = r^{\alpha - 1} f_1(\theta)
\cal{F}(\theta) \ , 
\label{eq: lenseqourb}
\end{equation}

where we have defined

\begin{equation}
f_1(\theta) = \frac{2\delta \beta \sin{(2\theta - 2\theta_p)}} {1 - \delta \cos{(2\theta - 2\theta_p)}} \ .
\label{eq: funodef}
\end{equation}

Having four images\footnote{In the following we add to image coordinates a lower index running on $i, j, k, l$ to
distinguish among them.}, in principle we may try to solve the eight equations obtained inserting image positions
in Eqs.(\ref{eq: lenseqoura}) and (\ref{eq: lenseqourb}) to determine the potential parameters $(\alpha, \beta,
\delta, \theta_p)$ and the source coordinates $(r_s, \theta_s)$. Unfortunately this is not possible neither
analytically nor numerically since, in these equations, $\alpha$ and $\beta$ enter as exponent and this causes the
system to be nonlinear and trascendent. However we will show, in the following, how it is possible to
algebrically manipulate these equations in such a way to get a more handable and numerically solvable system.

As a first step, let us insert the coordinates $(r_i, \theta_i)$ of image $i$ into Eqs.(\ref{eq: lenseqoura}) and
(\ref{eq: lenseqourb}) and divide side by side; one gets

\begin{equation}
\frac{r_i - r_s \cos{(\theta_i - \theta_s)}}{r_s \sin{(\theta_i - \theta_s)}} =
 \frac{\alpha [ 1 - \delta \cos{(2\theta_i - 2\theta_p)}]}{2\delta \beta \sin{(2\theta_i - 2\theta_p)}} \ . 
\label{eq: step}
\end{equation}

This relation does not hold if $\sin{(\theta_i - \theta_s)} = 0$ and/or $\sin{(2\theta_i - 2\theta_p)} = 0$.
Exploiting the lens equations (\ref{eq: lenseqoura}) and (\ref{eq: lenseqourb}), it is easy to show that these
divergences are very unlikely. Suppose that $\sin{(2\theta_i - 2\theta_p)} = 0$; in this case Eq.(\ref{eq:
lenseqourb}) reduces to

\begin{displaymath}
r_s \sin{(\theta_i - \theta_s)} = 0
\end{displaymath}

which is satisfied only if $r_s = 0$ or if $\sin{(\theta_i - \theta_s)} = 0$. The first solution ($r_s = 0$)
means that the source and the galaxy are perfectly aligned which is very unlikely so that we may discard this case. The other solution ($\sin{(\theta_i - \theta_s)} = 0$) is also very unlikely since it means that the source is aligned with one of the images. For these reasons we may assume that Eq.(\ref{eq: step}) holds for all images and never diverges. The same equation may be also written  for images $j, k, l$; dividing side by side the one for image $i$ by the ones for the other images, one gets the following three equations\,:

\begin{displaymath}
\frac{r_i - r_s \cos{(\theta_i - \theta_s)}}{r_j - r_s \cos{(\theta_j - \theta_s)}} \ \frac{\sin{(\theta_j -
\theta_s)}}{\sin{(\theta_i - \theta_s)}} = 
\end{displaymath}
\begin{equation}
\frac{ 1 - \delta \cos{(2\theta_i - 2\theta_p)}}{ 1 - \delta
\cos{(2\theta_j - 2\theta_p)}} \ \frac{\sin{(2\theta_j - 2\theta_p)}}{\sin{(2\theta_i - 2\theta_p)}} \ ,
\label{eq: suno}
\end{equation}

\begin{displaymath}
\frac{r_i - r_s \cos{(\theta_i - \theta_s)}}{r_k - r_s \cos{(\theta_k - \theta_s)}} \ \frac{\sin{(\theta_k -
\theta_s)}}{\sin{(\theta_i - \theta_s)}} = 
\end{displaymath}
\begin{equation}
\frac{ 1 - \delta \cos{(2\theta_i - 2\theta_p)}}{ 1 - \delta
\cos{(2\theta_k - 2\theta_p)}} \ \frac{\sin{(2\theta_k - 2\theta_p)}}{\sin{(2\theta_i - 2\theta_p)}} \ ,
\label{eq: sdue}
\end{equation}

\begin{displaymath}
\frac{r_i - r_s \cos{(\theta_i - \theta_s)}}{r_l - r_s \cos{(\theta_l - \theta_s)}} \ \frac{\sin{(\theta_l -
\theta_s)}}{\sin{(\theta_i - \theta_s)}} = 
\end{displaymath}
\begin{equation}
\frac{ 1 - \delta \cos{(2\theta_i - 2\theta_p)}}{ 1 - \delta
\cos{(2\theta_l - 2\theta_p)}} \ \frac{\sin{(2\theta_l - 2\theta_p)}}{\sin{(2\theta_i - 2\theta_p)}} \ .
\label{eq: stre}
\end{equation}

Note that $\alpha$ and $\beta$ never appear in Eqs.(\ref{eq: suno}), (\ref{eq: sdue}), (\ref{eq: stre}) which are
thus three independent equations for the four unknowns\,: $(r_s, \theta_s, \delta, \theta_p)$. The system is not
closed (i.e., the number of equations is not equal to the number of variables) and so cannot be solved unless we
add a further independent relation not involving $\alpha$ and $\beta$. To this aim, let us consider again the lens
equations for image $i$; solving the second one with respect to $r_i^{\alpha - 1} \cal{F}$\,$(\theta_i)$ and
substituting into the first one, it turns out

\begin{displaymath}
r_i - r_s \cos{(\theta_i - \theta_s)} = 
\end{displaymath}
\begin{displaymath}
\frac{\alpha [ 1 - \delta \cos{(2\theta_i - 2\theta_p)}]}{2\delta \beta
\sin{(2\theta_i - 2\theta_p)}} \ r_s \sin{(\theta_i - \theta_s)} \ .
\end{displaymath}

Writing down the same relation for the image $j$ and subtracting from the one for image $i$, one gets\,:

\begin{displaymath}
(r_i - r_j) - r_s [\cos{(\theta_i - \theta_s)} - \cos{(\theta_j - \theta_s)}] =
\end{displaymath}

\begin{displaymath}
\frac{r_s \alpha}{2 \delta \beta} \left \{ \frac{\sin{(\theta_i - \theta_s) [1 - \delta \cos{(2 \theta_i - 2
\theta_p )}]}} {\sin{(2 \theta_i - 2 \theta_p )}} -
\right .
\
\end{displaymath}
\begin{equation}
\ \ \ \ \ \ \ \
\left .
\ 
\frac{\sin{(\theta_j - \theta_s) [1 - \delta \cos{(2 \theta_j
- 2 \theta_p )}]}} {\sin{(2 \theta_j - 2 \theta_p )}} \right \} \ . 
\label{eq: presquattro}
\end{equation}

The same expression may be written also using images $k$ and $l$ instead of $i$ and $j$; dividing side by side
these relations, one finally finds out the following equation\,:

\begin{displaymath}
\frac{(r_i - r_j) - r_s [\cos{(\theta_i - \theta_s)} - \cos{(\theta_j - \theta_s)}]} {(r_k - r_l) - r_s
[\cos{(\theta_k - \theta_s)} - \cos{(\theta_l - \theta_s)}]} =
\end{displaymath}

\begin{displaymath}
\left \{ \frac{\sin{(\theta_i - \theta_s) [1 - \delta \cos{(2 \theta_i - 2 \theta_p )}]}} {\sin{(2 \theta_i - 2
\theta_p )}} - 
\right .
\ 
\end{displaymath}
\begin{displaymath}
\ \
\left .
\
\frac{\sin{(\theta_j - \theta_s) [1 - \delta \cos{(2 \theta_j - 2 \theta_p )}]}} {\sin{(2 \theta_j
- 2 \theta_p )}} \right \} \times
\end{displaymath}

\begin{displaymath}
\left \{ \frac{\sin{(\theta_k - \theta_s) [1 - \delta \cos{(2 \theta_k - 2 \theta_p )}]}} {\sin{(2 \theta_k - 2
\theta_p )}} - 
\right .
\
\end{displaymath}
\begin{equation} 
\ \
\left .
\
\frac{\sin{(\theta_l - \theta_s) [1 - \delta \cos{(2 \theta_l - 2 \theta_p )}]}} {\sin{(2 \theta_l
- 2 \theta_p )}} \right \}^{-1} \ . 
\label{eq: squattro}
\end{equation}

This equation does not contain neither $\alpha$ nor $\beta$ and thus can be added to Eqs.(\ref{eq: suno}),
(\ref{eq: sdue}) and (\ref{eq: stre}) to get a closed system of four independent equations in the four variables
$(r_s, \theta_s, \delta, \theta_p)$.

This system may be rewritten in a more convenient way. Actually, solving each of the Eqs.(\ref{eq: suno}), (\ref{eq: sdue}), (\ref{eq: stre}) and (\ref{eq: squattro}) with respect to $r_s$ we simply get\,:
\begin{equation}
r_s = \frac{\lambda_{ij} \delta + \mu_{ij}}{\nu_{ij} \delta + \eta_{ij}} \ ,
\label{eq: sfuno}
\end{equation}
\begin{equation}
r_s = \frac{\lambda_{ik} \delta + \mu_{ik}}{\nu_{ik} \delta + \eta_{ik}} \ ,
\label{eq: sfdue}
\end{equation}
\begin{equation}
r_s = \frac{\lambda_{il} \delta + \mu_{il}}{\nu_{il} \delta + \eta_{il}} \ ,
\label{eq: sftre}
\end{equation}
\begin{equation}
r_s = \frac{p_1 \delta + p_0}{q_1 \delta + q_0} \ ,
\label{eq: sfquattro}
\end{equation}
where the coefficients depend only on the two angles $(\theta_s, \theta_p)$ and are defined as follows\footnote{For sake of shortness we give the expressions only for some coefficients, the other ones being defined in a similar way changing $j$ with $k$ or $l$.}\,:
\begin{displaymath}
\lambda_{ij} = r_j \sin{(\theta_i - \theta_s)} \sin{(2\theta_j - 2\theta_p) \cos{(2\theta_i - 2\theta_p)}} 
\end{displaymath}
\begin{displaymath}
\ \ \ \ \ \ 
 - r_i \sin{(\theta_j - \theta_s)} \sin{(2\theta_i - 2\theta_p) \cos{(2\theta_j - 2\theta_p)}} \ ,
\end{displaymath}
\begin{displaymath}
\mu_{ij} = r_i \sin{(\theta_j - \theta_s)} \sin{(2\theta_i - 2\theta_p)} 
\end{displaymath}
\begin{displaymath}
\ \ \ \ \ \ 
-  r_j \sin{(\theta_i - \theta_s)} \sin{(2\theta_j - 2\theta_p)} \ ,
\end{displaymath}
\begin{displaymath}
\nu_{ij} = \cos{(\theta_j - \theta_s)} \sin{(\theta_i - \theta_s)} \sin{(2\theta_j - 2\theta_p)} \cos{(2\theta_i - 2\theta_p)}
\end{displaymath}
\begin{displaymath}
\ \ \ \ 
- \cos{(\theta_i - \theta_s)} \sin{(\theta_j - \theta_s)} \sin{(2\theta_i - 2\theta_p)} \cos{(2\theta_j - 2\theta_p)} \ ,
\end{displaymath}
\begin{displaymath}
\eta_{ij} = \cos{(\theta_i - \theta_s)} \sin{(\theta_j - \theta_s)} \sin{(2\theta_i - 2\theta_p)} 
\end{displaymath}
\begin{displaymath}
\ \ \ \ 
- \cos{(\theta_j - \theta_s)} \sin{(\theta_i - \theta_s)} \sin{(2\theta_j - 2\theta_p)} 
\end{displaymath}
\begin{displaymath}
p_1 = (r_i - r_j) \sigma_{kl} \sin{(2\theta_i - 2\theta_p) \sin{2\theta_j - 2\theta_p}} 
\end{displaymath}
\begin{displaymath}
\ \ \ \ 
- (r_k - r_l) \sigma_{ij} \sin{(2\theta_k - 2\theta_p) \sin{2\theta_l - 2\theta_p}} \ ,
\end{displaymath}
\begin{displaymath}
p_0 = (r_i - r_j) \tau_{kl} \sin{(2\theta_i - 2\theta_p) \sin{2\theta_j - 2\theta_p}} 
\end{displaymath}
\begin{displaymath}
\ \ \ \ 
- (r_k - r_l) \tau_{ij} \sin{(2\theta_k - 2\theta_p) \sin{2\theta_l - 2\theta_p}} \ ,
\end{displaymath}
having posed for sake of shortness\,:
\begin{displaymath}
\sigma_{ij} = \sin{(\theta_j - \theta_s)} \sin{(2\theta_i - 2\theta_p)} \cos{(2\theta_j - 2\theta_p)} 
\end{displaymath}
\begin{displaymath}
\ \ \ \ 
- \sin{(\theta_i - \theta_s)} \sin{(2\theta_j - 2\theta_p)} \cos{(2\theta_i - 2\theta_p)} \ ,
\end{displaymath}
\begin{displaymath}
\tau_{ij} = \sin{\theta_i - \theta_s} \sin{(2\theta_j - 2\theta_p)} - \sin{\theta_j - \theta_s} \sin{(2\theta_i - 2\theta_p)} \ .
\end{displaymath}
It is now very simple to reduce the system of Eqs.(\ref{eq: sfuno}), (\ref{eq: sfdue}), (\ref{eq: sftre}) and (\ref{eq: sfquattro}) to a system of three equations in only  three unkwons; equating side by side those equations one finally gets the following closed systems in $(\theta_s, \delta, \theta_p)$\,:
\begin{equation}
a_1 \delta^2 + b_1 \delta + c_1 = 0 \ ,
\label{eq: effeuno}
\end{equation} 
\begin{equation}
a_2 \delta^2 + b_2 \delta + c_2 = 0 \ ,
\label{eq: effedue}
\end{equation} 
\begin{equation}
a_3 \delta^2 + b_3 \delta + c_3 = 0 \ ,
\label{eq: effetre}
\end{equation} 
with the coefficients depending only on $(\theta_s, \theta_p)$ and defined as follows\,:
\begin{displaymath}
a_1 = \lambda_{ij} \nu_{ik} - \lambda_{ik} \nu_{ij} \ ,
\end{displaymath}
\begin{displaymath}
b_1 = \lambda_{ij} \eta_{ik} + \mu_{ij} \nu_{ik} -  \lambda_{ik} \eta_{ij} + \mu_{ik} \nu_{ij} \ ,
\end{displaymath}
\begin{displaymath}
c_1 = \mu_{ij} \eta_{ik} - \mu_{ik} \eta_{ij} 
\end{displaymath}
and similarly for $(a_2, b_2, c_2)$ (changing $k$ with $l$) and  $(a_3, b_3, c_3)$ (changing $\lambda_{ik}$, $\mu_{ik}$, $\nu_{ik}$, $\eta_{ik}$ with $p_1$, $p_0$, $q_1$, $q_0$). The system formed by Eqs.(\ref{eq: effeuno}), (\ref{eq: effedue}) and (\ref{eq: effetre}) is closed and may be solved numerically to find the three unknown parameters $(\theta_s, \delta, \theta_p)$. Then, one of the Eqs. (\ref{eq: sfuno}),  (\ref{eq: sfdue}),  (\ref{eq: sftre}) and (\ref{eq: sfquattro}) may be used to finally get the other source coordinate $r_s$. Actually, it is still possible to further reduce the system using the {\it resultant method} (\cite{ES93}) to get only two equations in $(\theta_s, \theta_p)$ which have to be solved numerically. However we have preferred to not apply this method since it is difficult to implement in a totally automathic code and it leads to no significative decrease of the CPU time with respect to the strategy of solving directly Eqs. (\ref{eq: effeuno}), (\ref{eq: effedue}) and (\ref{eq: effetre}). Beside, some tests have shown us that the two approaches lead to the same results.  

Having found the four parameters $(r_s, \theta_s, \delta, \theta_p)$, it is now straightforward to find the other two lensing potential parameters $\alpha$ and $\beta$. To this aim let us consider again Eq.(\ref{eq: lenseqoura}) written for images $i$ and $j$; dividing side by side one gets\,:

\begin{displaymath}
\frac{r_i - r_s \cos{(\theta_i - \theta_s)}}{r_j - r_s \cos{(\theta_j - \theta_s)}} = 
\end{displaymath}
\begin{equation}
\left ( \frac{r_i}{r_j}
\right )^{\alpha - 1} \left [ \frac{1 - \delta \cos{(2\theta_i - 2\theta_p)}}{1 - \delta \cos{(2\theta_j -
2\theta_p)}} \right ]^{\beta}  \ . 
\label{eq: alfabeta}
\end{equation}

The same holds for images $k$ and $l$. Taking the natural logarithm of these expressions and solving for $\alpha$
and $\beta$ one finally finds\,:

\begin{equation}
\alpha = \frac{a_{kl} b_{ij} - a_{ij} b_{kl}}{b_{ij} c_{kl} - b_{kl} c_{ij}} + 1 \ , 
\label{eq: alfa}
\end{equation}

\begin{equation}
\beta = \frac{a_{ij} - c_{ij} (\alpha - 1)}{b_{ij}} \ , 
\label{eq: beta}
\end{equation}

where we have defined\,:

\begin{equation}
a_{ij} = \ln{\left [ \frac{r_i - r_s \cos{(\theta_i - \theta_s)}}{r_j - r_s \cos{(\theta_j - \theta_s)}} 
\right ]} \ , 
\label{eq: aij}
\end{equation}

\begin{equation}
b_{ij} = \ln{\left [\frac{1 - \delta \cos{(2\theta_i - 2\theta_p)}}{1 - \delta \cos{(2\theta_j - 2\theta_p)}}
\right ]} \ , 
\label{eq: bij}
\end{equation}

\begin{equation}
c_{ij} = \ln{(r_i/r_j)} \label{eq: cij} ,
\end{equation}

and similar relation for $a_{kl}$, $b_{kl}$, $c_{kl}$. Note that all the quantities entering Eqs.(\ref{eq: alfa})
and (\ref{eq: beta}) are known since image positions are observed and $(r_s, \theta_s, \delta, \theta_p)$ have
been found solving the system formed by Eqs.(\ref{eq: sfuno}), (\ref{eq: effeuno}), (\ref{eq: effedue}) and (\ref{eq: effetre}). These latter equations, Eqs.(\ref{eq: alfa}) and (\ref{eq: beta}) allow us to find out the source
positions and the lensing potential parameters by knowing only the images positions and without using any
information on the flux ratios (which may be affected by microlensing, that is images cannot be resolved) and/or
the time delays among images (which are measured only for two quadruple systems).

\section{Solving numerically for potential parameters}

Eqs.(\ref{eq: effeuno}), (\ref{eq: effedue}), (\ref{eq: effetre}) are a system of three equations
which may be solved to find out the source position angle $\theta_s$,  the flattening indicator $\delta$ and the position angle $\theta_p$ of the lensing potential. Then Eq.(\ref{eq: sfuno}) gives the other source coordinate $r_s$ and, finally, Eqs.(\ref{eq: alfa}) and (\ref{eq: beta}) allow to get the other two potential parameters, the slope $\alpha$ of the radial profile and the boxiness parameter $\beta$. The system formed by the first three equations is not linear and its solutions may be found only numerically. This has led us to develop an algorithm\footnote{The code is nothing more but a notebook written for MATHEMATICA, which we have named {\it PULP} ({\it Probing Unknown Lens Parameters}).} to search for the solutions of the system. To avoid introducing any bias in the search, we give to the algorithm $\cal{N}$ random starting points for $(\theta_s, \delta, \theta_p)$, where $\cal{N}$ is a number fixed by
the user\footnote{$\cal{N}$ should be large to explore a wide region in the parameter space, but not too large to
save computer time. The right choice must be a compromise between these two different circumstances. A possible
strategy could be to fix $\cal{N} =$\,4000 and then, eventually, run PULP more than one time if necessary.}. We
then obtain $\cal{N}$ solutions which are not all different from each other and are not all physically
acceptable. To select among these we have imposed a set of selection criteria\,: the code checks the list of
$\cal{N}$ solutions and finally retains only the ones satisfying the whole set of criteria. Schematically the
selection procedure works as follows\,:

\begin{enumerate}

\item{check that the solutions $(\theta_s, \delta, \theta_p)$ are good approximations of the real ones, i.e. retain only those solutions which inserted again into Eqs.(\ref{eq: effeuno}), (\ref{eq: effedue}) and (\ref{eq: effetre}) solve them with high accuracy;}
\item{evaluate $r_s$ from Eqs.(\ref{eq: sfuno}), (\ref{eq: sfdue}), (\ref{eq: sftre}), (\ref{eq: sfquattro}) and exclude all solutions which give values of $r_s$ different among each other and/or negative\,: this test is necessary to be sure that the quadruplet $(r_s, \theta_s, \delta, \theta_p)$ indeed solve the system (i.e. it is not a false solution due to a bad convergence of the numerical code) and to erase the unphysical solutions with negative $r_s$;}
\item{select solutions with $\delta < 1$\,: this is necessary to be consistent with the starting hypotheses;}

\item{look for solutions with $\alpha < 2$ in order to have $\Sigma(r, \theta)$ monotonically decreasing with the radius;}

\item{exclude all solutions with $| \beta | > 1/2$ since these often lead to unphysical mass\,-\,radius relation (\cite{ZP00});}

\item{finally retain only solutions which give rise to galactic models with $0.2 \le q_{\kappa} \le 1.0$ where $q_{\kappa}$ is evaluated through Eq.(\ref{eq: qkappa}); this latter constraint has been added since galactic halos (which represent more than 90\% of the lens galaxies mass) are never prolate and their flattening (even if not well constrained) is always larger than 0.4 so that 0.2 is a quite conservative lower limit.}

\end{enumerate}

Applying this set of constraints eliminates a lot of solutions leaving us with only $\cal{M}$ solutions, being
$\cal{M}$ a number much lower than $\cal{N}$. There are now two possibilities. The first is $\cal{M} =$\,0; this
may be due to a wrong choice of the potential, that is the lensing potential does not belong to the class
described by Eq.(\ref{eq: psiour}), but to be sure that this is indeed the case, it is better to run the code
more than one time with $\cal{N}$ very large in order to explore a very wide region of the parameter space. The
other hypothesis is that $\cal{M} >$\,1, i.e. we still have more than one solution\footnote{Unfortunately the
case $\cal{M} =$\,1 is very unlikely.}. After having checked that these $\cal{M}$ solutions are really different
(since it is still possible that one has the same solution except for the values of $\theta_s$ and/or $\theta_p$
which, being angles, are defined $mod \ 2\pi$), we have to further select among these ones. A powerful
discriminator is the time delay ratio among different couples of images, but unfortunately this is not useful in
practice since only two of the fifteen observed quadruple systems have measured time delays\footnote{Detailed
informations on the observed lens systems, both double and quadruple, may be found in the CASTLES collaboration
web page (\cite{CASTLE}).}. On the other hand flux ratios are always measured with a good accuracy; these quantities depend on the lensing potential parameters, so that one may use their observed values as constraints to select among the $\cal{M}$ solutions previously found.

To this aim, let us first evaluate the flux ratios for the class of lensing potentials described by Eq.(\ref{eq:
psiour}). Let $\phi_s$ be the (unknown) source flux and $\phi_i$ the one coming from the image with coordinates
$(r_i, \theta_i)$. The magnification due to the lensing effect is given by (\cite{SEF})\,:

\begin{equation}
\mu(r_i, \theta_i) = \frac{\phi_i}{\phi_s} = | (1 - \kappa )^2 - \gamma_1^2 - \gamma_2^2 |^{-1} \ , 
\label{eq: defmu}
\end{equation}

being\,:

\begin{equation}
\kappa = \frac{1}{2} \left ( \frac{\partial^2}{\partial x^2} + \frac{\partial^2}{\partial y^2} \right ) \psi \ ,
\label{eq: kappadef}
\end{equation}

\begin{equation}
\gamma_1 = \frac{1}{2} \left ( \frac{\partial^2}{\partial x^2} - \frac{\partial^2}{\partial y^2} \right ) \psi \ ,
\label{eq: gammaunodef}
\end{equation}

\begin{equation}
\gamma_2 = \frac{1}{2} \frac{\partial^2 \psi}{\partial x \partial y} \ , \label{eq: gammaduedef}
\end{equation}

which are, respectively, the dimensionless surface mass density, yet introduced in Eq.(\ref{eq: psikappa}), and
the two components of the shear vector (\cite{SEF}).

Introducing Eq.(\ref{eq: psiour}) into Eqs.(\ref{eq: kappadef}), (\ref{eq: gammaunodef}), (\ref{eq: gammaduedef})
and then the results into Eq.(\ref{eq: defmu}), after a lengthy calculation (being careful that our
transformation to polar coordinates is a little bit different from the usual) one finally gets the magnification
which is

\begin{displaymath}
\mu^{-1}(r, \theta) = | 1 - (\alpha^2 + f_1^2 + f_2) r_i^{\alpha - 2} \cal{F}(\theta) \ +
\end{displaymath}

\begin{equation}
\; \; \; \; \; \; \; \; (\alpha - 1) \ (\alpha^2 + f_1^2 + \alpha f_2) r_i^{2(\alpha - 2)} \cal{F}(\theta) |
\label{eq: muour}
\end{equation}

being\,:

\begin{equation}
f_2(\theta) = \frac{4 \delta \beta [\cos{}(2\theta - 2\theta_p) - \delta]} {[1 - \delta \cos{(2\theta -
2\theta_p)}]^2} \ . 
\label{eq: fdue}
\end{equation}

The flux ratios between images $i$ and $j$ is then simply\,:

\begin{equation}
\phi_{ij} = \frac{\phi_j}{\phi_i} = \frac{\phi_j}{\phi_s} \times \frac{\phi_s}{\phi_i} = \frac{\mu(r_j,
\theta_j)}{\mu(r_i, \theta_i)} \label{eq: fluxijay}
\end{equation}

which is more convenient to express in magnitudes as\,:

\begin{equation}
m_{ij} = m_j - m_i = -2.5 \log{\phi_{ij}} \ ; \label{eq: mijay}
\end{equation}

being $m_i$ $(m_j)$ the observed magnitudes of image $i$ $(j)$.

For each solution among the $\cal{M}$ previously selected, we may now evaluate the flux ratios with respect to the
image $i$ and compare these with the observed ones finally retaining only those solutions such that

\begin{equation}
\left | \frac{m_{ij}(obsd) - m_{ij}(sol)}{m_{ij}(obsd)} \right | \le \varepsilon ,\label{eq: criteriamij}
\end{equation}

being $m_{ij}(obsd)$ the observed value, $m_{ij}(sol)$ the ones predicted by the examined solution and
$\varepsilon$ a tolerance fixed by the user. The value of $\varepsilon$ is difficult to choose for two main
reasons. On the one hand, $m_{ij}$ is extremely sensitive to the values of $\alpha$ and $\beta$ since these
quantities enter as exponents in the theoretical formula. That is why these parameters must be finely tuned to be
in agreement with observations since a small error on their values may lead to a high error on $m_{ij}$. On the
other hand, flux ratios may be affected by microlensing which may change the flux of only one of the images
leading to a flux ratio which is different from what is predicted using Eqs.(\ref{eq: muour}) and (\ref{eq:
fluxijay}). These reasons lead us to not choose a too small value of $\varepsilon$ in order to avoid to exclude
solutions not satisfying the constraint (\ref{eq: criteriamij}), which are indeed good approximations of the right
one. Some tests have shown that a good choice could be $\varepsilon = 0.5$ even if a so high value is not too restrictive. Given this problems of fine tuning, we have decided to not use the flux ratios as constraints when modelling real lenses since it is very difficult to reach a so high accuracy in recovering the parameters $\alpha$ and $\beta$ when working with data affected by observational errors.

The algorithm searches for solutions and matches  the whole set of constraints finally leading to a very small
number of solutions or directly selecting only one solution. In order to be sure that this is the correct
solution one may also recalculate the image positions to see if they are in agreement (within the observational
errors) with the observed ones and finally choose only one solution.

\subsection{A test example}

To test if our code works well or not, i.e. if it is able to recover the correct solution solving the system given
by Eqs.(\ref{eq: sfuno}), (\ref{eq: effeuno}), (\ref{eq: effedue}) and (\ref{eq: effetre}), we have made some tests on simulated situations. As an example,  we describe  one of these.

As a first step, we have given the source position $(r_s, \theta_s)$ as $(0.09'', 35^o)$ and chosen the lensing
potential parameters as $(\alpha, \beta, \delta, \theta_p) = (1.3, -0.25, -0.25, 60^o.5)$. Then, we have
numerically solved lens equations (\ref{eq: lenseqoura}) and (\ref{eq: lenseqourb}) for this potential
determining image positions. We have also evaluated flux ratios with respect to image $i$ and finally we have
approximated all these quantities to simulate  the  image positions\footnote{The whole task has been done with a
 MATHEMATICA notebook written by us and named {\it LIGeIA} ({\it Lensing Images Generator Implemented
Algorithm}). Both LIGeIA and PULP are available on request to the authors.}. These quantities are then used as
input for the algorithm; we have then fixed $\cal{N} =$\,1000 and then started the search. The first set of
selection criteria (not considering the one on the flattening) gives 72 ($\sim 7\%$ of the starting ones)
solutions, which further reduces to  24 ($\sim 2\%$ of the starting ones) after the constrain on $q_{\kappa}$.
Finally the conservative constraint on the flux ratios leaves us with only one solution\,:

\begin{displaymath}
(r_s, \theta_s, \delta, \theta_p, \beta, \alpha) = 
\end{displaymath}
\begin{displaymath}
(0.09'', 34^o.6, -0.24, 240^o.5, -0.26, 1.29)
\end{displaymath}

which is a very good approximation of the starting parameters given in the simulation. Note, however, that the
recovered value of $\theta_p$ differs from the real one of $\pi$, which is however a well known degeneracy which we do not comment on further.  

The test has been repeated many times with different sets of source positions and potential parameters and has
shown that our code always recovers the exact values of the parameters with a quite good accuracy.

\section{Application to PG1115+080 and estimate of the Hubble constant}

Having checked that the semi\,-\,analytical method indeed works in recovering the lensing potential parameters, the next step is to apply it to real lenses to see whether the the results are in agreement with previous ones in literature. To this aim, we need a system with four images and a good astrometry not only of the image positions, but also of the lens galaxy centre since this latter will be used as the origin of our coordinate system. Beside, we also need that there is only one galaxy acting as lens otherwise it is not possible to factorize the lensing potential. In the CASTLES database (\cite{CASTLE}) there are fifteen quadruple lenses but only some of them satisfy (more or less) all our requirements. Among these we have chosen to consider the well known PG1115+080, first discovered by Weymann et al. (1980) and then studied in detail by several authors (see, e.g, \cite{KK97}; \cite{WS00}; \cite{ZP00}) with different techniques. This system consists of four images (named $A1$, $A2$, $B$ and $C$) of a radio quiet QSO at $z_s = 1.722$, while the lens is an elliptical galaxy belonging to a group of $\sim$\,10 galaxies at $z_L = 0.310$. The center of the group is at $(r_g, \theta_g) = (20'' \pm 0.2'', -117^o \pm 3^o)$ and its effect has been taken into account as an external shear in previous models. This consideration should suggest that PG1115+080 is not the best system to apply our method since we have neglected any contribution of the external shear from the beginning. However, we note that previous models never considered boxy potentials\footnote{Zhao \& Pronk (2001) indeed used also potentials similar to the our one, but in their method the boxiness parameter is fixed by hand from the beginning.} as we do and so it is interesting to explore also this possibility. Beside, PG1115+080 is one of the two quadruple lenses with measured time delays\footnote{The other one is B1608+656 which is too difficult to model since there are two lensing galaxies which are probably undergoing a merging event.} which allows us, on the one hand, to use also the time delay ratio to select among the possible solutions and, on the other hand, to get an estimate of the Hubble constant $H_0$.         
In the following subsections we apply the method to this system using as images coordinates the ones measured by Impey et al. (1998) with HST observations; the time delay between images $A1$ and $A2$ is too small to be measured, while the one between images $B$ and $C$ is $\Delta t_{BC} = 25.0 \pm 1.7$\,d (\cite{Barkana}) with image $B$ arriving last. There are also two different measures of the time delay ratio among images $BC$ and $AB$\,: Schechter et al. (1997) first reported $r_{ABC} = \Delta t_{AB}/\Delta t_{BC} = 0.7 \pm 0.3$, while a later analysis by Barkana (1997) found $r_{ABC} = 1.13 \pm  0.18$. First, we do some general considerations on how to treat the errors. 

\subsection{Effect of observational errors}

Our semi\,-\,analytical method has been tested on simulated cases implicitly assuming that the image positions are measured with no errors. However, when working with real data, one has to take into account also the errors and how these affect the determination of the lensing potential parameters. Since we do not have anay analytical formula which gives the solutions as function of the observed image coordinates, it is not possible to propagate the errors from image positions to lens parameters. However, it is still possible to give a qualitative (but quite conservative) estimate of these latter quantities. To this aim, we have added a $5\%$ uncertainty on the image coordinates of one of our simulated systems; then, we have run the code more than 100 times each time given randomly the image coordinates each one within the range obtained adding the error. Each one of these run has lead us to an estimate of the source coordinates $(r_s, \theta_s)$ and of the potential parameters $(\alpha, \beta, \delta, \theta_p)$ which we may compare to the real values of the parameters which we have assigned from the beginning. These tests have shown us that the code still recovers the correct values of the parameters within an error which is less than 15\%. To be conservative we will add a 20\% error on the reconstructed parameter when working with real systems. 

This error on source coordinates and potential parameters induce an error also on the recalculated image positions which we have to estimate. To this aim, we adopt a similar procedure solving the lens equations (\ref{eq: lenseqa}), (\ref{eq: lenseqb}) for over 100 realizations of the set of parameters, randomly choosing each one in the range $(0.8 p, 1.2 p)$, being $p$ one of the parameters. Then, to be conservative, we give as our estimate of the image positions the median of the values obtained and as range the ones which contains the 90\% of the obtained values. Even if not statistically well motivated, this procedure should be conservative enough to assure us that we are not underestimating the errors. 

\subsection{Modelling PG1115+080 and estimates of $H_0$}

Here we apply our method to PG1115+080; as first step, we estimate the coordinates $(r, \theta)$ of the four images from the one in Impey et al. (1998) translating the origin of the coordinate system from the centre of the $C$ image to the galaxy one. A summary of the observed quantities is reported in Table 1 where we report also the values predicted by our best model.

\begin{table}
\begin{center}
\begin{tabular}{|c|c|c|} 
\hline
Image & $(r, \theta)_{observed}$ & $(r, \theta)_{predicted}$ \\
\hline
A1  &  $(1.173 \pm 0.002, 258.78 \pm 0.10)$ & $(1.19_{-0.16}^{+0.26}, 250_{-29}^{+27})$ \\
A2  &  $(1.120 \pm 0.002, 283.05 \pm 0.10)$ & $(1.10_{-0.14}^{+0.19}, 288_{-26}^{+32})$ \\
B   &  $(0.950 \pm 0.002, 155.43 \pm 0.12)$ & $(0.93_{-0.09}^{+0.10}, 150_{-23}^{+24})$ \\
C   &  $(1.397 \pm 0.002, 40.86 \pm 0.09)$ & $(1.36_{-0.16}^{+0.21}, 43_{-24}^{+23})$ \\
\hline
\end{tabular}
\end{center}
\caption{Observed and predicted quantities for PG1115+080. The radial coordinate is in arcsec and the angular one is in degrees.}
\end{table} 
   
This latter has been obtained applying our method adopting as input the central values of the range determined for each image position. This gives us the solution with the following parameters\,:
\begin{displaymath}
(r_s, \theta_s) \simeq (0.11'', 351^o) \ ,
\end{displaymath}
\begin{displaymath} 
(\alpha, \beta, \delta, \theta_p) \simeq (1.12, -0.31, -0.31, 325^o) \ . 
\end{displaymath}
To these values we add a 20\% error for the reasons explained in the previous subsection. Now we have to recalculate the image positions to be sure that this set of parameters predicts the correct configuration of the images and no other visible images. To this aim we proceed as described before generating over 200 set of parameters $(r_s, \theta_s, \alpha, \beta, \delta, \theta_p)$ with each value chosen randomly in a range defined as $(0.8 r_s, 1.2 r_s)$ and similar for the other parameters. For each of this realizations we compute the flattening $q_{\kappa}$ of the surface mass density using Eq.(\ref{eq: qkappa}), the Hubble constant $h$ and the time delay ratio $r_{ABC}$. These two latter quantities may be estimated as follows\,:
\begin{displaymath}
h = \frac{\Delta t_{BC}^{-1} \tau_{100}}{ 2 \alpha} \{ (2 - \alpha)(r_C^2 - r_B^2)   
\end{displaymath}
\begin{equation}
\ \ \
-2 (1 -\alpha) r_s [ r_C \cos{(\theta_C - \theta_s)} - r_B \cos{(\theta_B - \theta_s)} ] \} \ ,
\label{eq: hest}
\end{equation}
\begin{displaymath}
r_{ABC} = \{ (2 - \alpha)(r_C^2 - r_B^2) -2 (1 -\alpha) r_s \ \times 
\end{displaymath}
\begin{displaymath}
\ \ \ \ \ \ \ \ \ \ \ \ 
[ r_C \cos{(\theta_C - \theta_s)} - r_B \cos{(\theta_B - \theta_s)} ] \}
\ \times
\end{displaymath}
\begin{displaymath}
\{ (2 - \alpha)(r_{a1}^2 - r_B^2) -2 (1 -\alpha) r_s \ \times 
\end{displaymath}
\begin{equation} 
[ r_{A1} \cos{(\theta_{A1} - \theta_s)} - r_B \cos{(\theta_B - \theta_s)} ] \}^{-1} \ .
\label{eq: ratiotime}
\end{equation}
In computing $h$ we adopt a flat cosmology with $\Omega_m = 0.3$ and $\Omega_{\Lambda} = 0.7$ which gives us $\tau_{100} = 33.37 \ {\rm d} \ {\rm arcsec^{-2}}$. Different cosmological parameters change the estimate of $h$ less than 3\% so that we can neglect their effect. From the list so obtained we erase all the set of parameters which generates models with $q_{\kappa}$ outside the range (0.2, 1.0) or a value of $h$ outside the range (0.1,1.0) or a negative $r_{ABC}$. This leaves us with a consistent set of models each one with its own four image positions. The final estimate of the coordinates are reported in Table 1 where the central value is the median of the distribution and the errors are defined such that the range so delineated contains the 90\% of the values. Similarly we estimate the flattening $q_{\kappa}$ and the Hubble constant which turn out to be\,:
\begin{displaymath}
q_{\kappa} = 0.52_{-0.10}^{+0.07} \ \ \ \
H_0 = 56_{-11}^{+12} \ {\rm km \ s^{-1} \ Mpc^{-1}} \ .
\end{displaymath}
The mean value of $r_{ABC}$ turns out to be 0.79, but the values are so sparse that defining a 90\% range is meaningless. Note, however, that this value is in agreement with the one obtained by Schechter et al. (1997). We deserve to conclusions the discussion of these results. Finally, note that we have also found other solutions, but we have deleted them since the predicted time delay ratio is grossly inconsistent with the estimates of both Schetchter et al. (1997) and Barkana (1997).

\section{Discussion and Conclusions}

In this paper we have developed a semi\,-\,analytical method to reconstruct lensing potential in quadruply imaged
systems. The method relies only on the image positions and allows to recover both the source coordinates and the
whole set of potential parameters for a broad class of non\,-\,elliptical boxy potentials. The angular structure of the lens potential is a crucial feature of the lens models, so it is important to generalize from elliptical to boxy lens models to understand how the lens models are affected by boxiness. Boxy potentials similar to our models have been yet considered by Zhao \& Pronk (2001), but our method is the first which allows to recover all the model parameters without giving {\it a priori} the boxiness parameter $\beta$.  The method does not use time delays ratios between any two images since these very useful quantities are so difficult to measure that nowadays only two quadruple systems have been used to measure time delays. The flux ratios are used only as broad constraints  to avoid the problems connected to the possible systematic effect of microlensing; however, they turn out to be very sensitive to the slope $\alpha$ of the radial profile and the boxiness parameter $\beta$ so that a fine tuning of these two quantities seems to be necessary to fit the flux ratios. The method has been tested on simulated cases to see if it works well or not, i.e. if it is able to correctly recover the potential parameters. This test has shown that it  works indeed being only affected by the well known degeneracy on the position angle $\theta_p$ (i.e. both $\theta_p$ and $\theta_p + \pi$ are acceptable solutions). To take into account the observational errors in the measurements of image coordinates, we have used a conservative approach to estimate the uncertainties induced on the recovered parameters and how they propagate on the image positions predicted by the reconstructed potential. Even if not statistically well motivated, our error estimates are quite conservative such that we are confident that we have not introduced any systematic errors in our reconstruction.

The method described has been applied to the quadruple lens PG1115+080; for this system we have also used the contraints coming from the measured time delay ratio and estimated the Hubble constant. Our best model fits well the four images positions which is a nice result especially if one considers that we have not introduced any external shear from the group the lens galaxy belongs to, as is usually done in the previous analyses of this system. This has allowed us also to recover an estimate of the Hubble constant, finding out $H_0 = 56_{-11}^{+12} \ {\rm km \ s^{-1} \ Mpc^{-1}}$. PG115+080 has been studied in detail by different authors using different techniques so that it is interesting to compare our result with the previous ones in literature. Keeton \& Kochanek (1997) have used the usual least $\chi^2$ parametric approach modelling the lensing potential as the sum of a term due to the galaxy and an external shear from the group. They have considered various elliptical models and have found that none of them may fit the image positions without any external shear. One of their two best fit models describes the lens galaxy as an isothermal (i.e. $\alpha = 1.0$) one with axis ratio $q = 0.90$ treating the group as a point mass. Our best fit model is nearly isothermal (having $\alpha = 1.12$), but the axis ratio is significantly different (being $q_{\kappa} = 0.52_{-0.10}^{+0.07}$). However, a direct comparison of the angular profile of the two models is not possible since the model by Keeton \& Kochanek is an elliptical one so that the its boxiness parameter is fixed to be $\beta = 0.5$, whilst our best fit model have $\beta = -0.31$ which is quite different. Actually, Keeton \& Kochanek start from lens mass models which are suggested by the light profile of the galaxy. Our approach aims at reconstructing the lensing potential; the result cannot be immediately compared to the light profile since the potential may be also generated by a dark halo embedding the visible component of the lens galaxy. It is thus difficult to say if the boxiness of the potential and the axis ratio of the surface mass density refer to the dark or visible component which lead us to not state any definitive conclusion from the comparison with Keeton \& Kochanek model. It is however interesting to compare the results on the Hubble constant with the previous ones in literature since a net discrepancy could suggest that something is wrong with our method. A carefull analysis of the possible systematic effects and of the different models suited for PG1115+080 finally lead Keeton \& Kochanek to estimate the Hubble constant as $H_0 = 51_{-13}^{+14} \ {\rm km \ s^{-1} \ Mpc^{-1}}$. This value is in quite good agreement with the our one, $H_0 = 56_{-11}^{+12} \ {\rm km \ s^{-1} \ Mpc^{-1}}$, which is a very encouraging result since we have used no contribution from external shear. Zhao \& Pronk (2001) have fitted a similar boxy model to describe PG1115+080, but a direct comparison with their result is not possible since they do not quote neither $\alpha$ nor $\delta$, but only what they called the {\it effective slope of the radial profile} and {\it a typical flattening}. Beside, they fix {\it a priori} the boxiness parameter, while in our method this is a reconstructed parameter. The disagreement between our and their values of the source coordinates may so be a consequence of the different boxiness parameters; however, we note that we predict a value for $H_0$ which is in agreement with what is reported in the second column of their Table 2. Finally, it is interesting to compare our result to the one obtained by Williams \& Saha (2000) using the fully non parametric {\it pixelated lens method}. In this approach, one does not try to reconstruct the lensing potential, but only the time arrival surface which, as they show, it is enough to estimate the Hubble constant. Combining the results from both PG1115+080 and B1608+656 they finally quote  $H_0 = 61 \pm 18 \ {\rm km \ s^{-1} \ Mpc^{-1}}$ (90\% confidence limit), still in agreement with what we have obtained.  

All these results are encouraging and seem to suggest that boxy potentials may describe quadruple lenses without the need of introducing an external shear. It is well known that this latter term is often necessary to fit well the image positions in quadruple systems and, beside, Witt (1996) has analytically shown that the minimum amount of shear, necessary to fit quadruple systems, is not trivial if the lensing potential is elliptical. However, we stress again that our models are not elliptical so that the result obtained by Witt may not be applied. Actually, we succeed in fitting the image positions in PG1115+080 also without introducing the external shear and have also predicted a value for the Hubble constant which is in good agreement with the previous estimates in literature obtained from this lens with different methods. These results are encouraging and we are now carrying on a detailed study of other quadruple lenses to see whether boxy potentials may fit the data and whether the predicted galaxy models are physically motivated (Cardone et al., 2001). On the other hand, it could be also necessary to consider boxy potentials with the external shear since many lens galaxies belong to a cluster whose effect should be considered before drawing conclusive results. This consideration lead us to further improve our method to take into account also the external shear. This task is not simple since to add the shear to the potential introduces other two variables, the shear strength $\gamma$ and its position angle $\theta_{\gamma}$, so that we will have eight variables. Image positions give us eight equations so that one may still hope to solve the problem since it is still possible to write a closed system of equations which we are tryng to manage to reduce to a numerically solvable one.

\begin{acknowledgements}
It is a pleasure to thank Lucia Mona for having helped us to solve some software related problems and Valerio Bozza for the interesting discussions we have had on the manuscript. We also warmly thank an anonymous referee whose comments have helped us to greatly improve the paper.
\end{acknowledgements}

\end{document}